\documentstyle{aipproc}
\input epsfig.sty
\def\gr{$\gamma$-ray }

\begin{document}
\title{The pulsar contribution to the diffuse galactic $\gamma$-ray emission}
\author{M. Pohl$^{1,2}$, G. Kanbach$^1$, S.D. Hunter$^3$, B.B. Jones$^4$}
\address{1 MPI f\"ur Extraterrestrische Physik, Postfach 1603, 85740 Garching,
Germany; \\
2 Danish Space Research Institute, Juliane Maries Vej 30, 2100 K\o benhavn \O,
Denmark; \\
3 LHEA, GSFC, Code 662, NASA, Greenbelt, MD 20771; \\
4 W.W. Hansen Laboratory, Dept. of Physics, Stanford University,
Stanford, CA 94305}

\maketitle
\begin{abstract}
Here we investigate to what extent
unresolved \gr pulsars contribute to the galactic diffuse emission, and
further whether unresolved \gr pulsars can be
made responsible for the excess of diffuse galactic emission above
1 GeV which has been observed by EGRET.
Our analysis is based only on the properties of the six pulsars which have
been identified in the EGRET data, and is independent of choice of a pulsar
emission model. 

We find that pulsars contribute very little to the diffuse emission at lower
energies, whereas above 1 GeV they can account for 25\%
of the observed intensity in selected regions for a reasonable number of 
directly observable \gr pulsars ($\sim$12).
While the excess above 1 GeV \gr energy
is observed at least up to six or eight degrees off the plane, 
the pulsar contribution would be negligible there. Thus pulsars do
significantly contribute to the diffuse galactic $\gamma$-ray emission
above 1 GeV, but they can not be made responsible for all the discrepancy
between observed intensity and model predictions in this energy range.
\end{abstract}

A recent analysis of the diffuse galactic \gr emission in the energy range
of 30 MeV to 30 GeV \cite{hunt97} indicated
that the spatial structure and total intensity of the emission observed
by EGRET can be well understood as the result of interactions between
cosmic rays and the interstellar medium in addition to a isotropic
extragalactic background. However, at energies above 1 GeV the models
predict only roughly 60\%
of the observed intensity.

The attempt of this paper is twofold. We want to provide constraints on
the general contribution of the most likely input from
discrete sources -- pulsars --
to the diffuse galactic \gr emission. We also want to find out whether or not
unresolved pulsars can account for the observed excess at high \gr energies.
Instead of using models as was done in previous studies
\cite{bk92,yadi95,sd96}
we will base our analysis solely on the properties of the 
six pulsars observed by EGRET.
A detailed description of the method can be found elsewhere \cite{po97}.

EGRET has up to now identified six pulsars by their light curves. The
differential \gr photon spectra of these sources have been derived
from pulsed analysis \cite{fier95}. We will concentrate on nine energy bands
spanning 50 MeV to 10 GeV, for which we have the observed photon flux
$S_i (E_k)$ of pulsar $i$ in energy band $E_k$. The basis
of our modelling is the assumption that pulsars do not behave
arbitrarily in their relation between the distance-normalized \gr
intensity and the age, but that they follow a trend, a correlation between
intensity and age. The
best evidence comes from the data themselves: the $\chi^2$ sums we obtain
indicate that a correlation is an appropriate description of the
data. The limited number of degrees
of freedom forces us to use the most simple correlation model, a
power-law:

\begin{equation}
S_m(t,E_k) = 10^{y_k}\, \left({{t}\over {10^4\ {\rm years}}}\right)^{b_k}
\quad {\rm kpc^2/ cm^2 / sec / MeV}
\end{equation}
This relation
can be fitted to the data 
on the basis of weighted least squares for all energy bins.
The distance $D_i$ and its uncertainty can be generally derived
on the basis of the pulsar dispersion
and rotation measure \cite{tayl93} and only for extremely
nearby objects like Geminga by parallax measurements \cite{cara96}.
The age of pulsars can be estimated from the ratio of period and period
derivative. The uncertainty of this measure of age is large, especially since
pulsars often do not slow down solely by dipole radiation. We will use a 
factor of two for the age uncertainty $\delta t_i$, except for the
Crab for which the true age is taken with a nominal uncertainty of 10\%.

Three sources of uncertainty have to be considered in the fit: uncertainties
in intensity, in distance, and in age. While the intensity error can be 
assumed to follow a Gaussian probability distribution, both the errors
in age and distance enter with some power, so that their effective probability
distributions are definitely not Gaussian. Furthermore, the way how
age and distance estimates are derived lets us think that a Gaussian
probability function is not a fair description of the actual error
distribution of age and distance, respectively.
We can account for these effects and still use the $\chi^2$ method,
if we assume that the error distribution for age and distance
are Gaussians in the logarithm, i.e. the uncertainty is multiplicative
rather than additive. Only in this case does any power of these parameters
obey the same corresponding error distribution.

To account for the different uncertainty distributions in intensity
on one side and age and distance on the other side,
we have separated each argument $\chi_i$ of the $\chi^2$-summation
into three components
$\chi_i^2 =\left(\chi_{i,1}^{-2}+\chi_{i,2}^{-2}+\chi_{i,3}^{-2}\right)^{-1}$,
one for
each source of uncertainty, where the components can be calculated
in the system in which the uncertainty distribution may be taken as Gaussian.
The best fit values and uncertainties of the parameters $y_k$ and $b_k$ of Eq.1 
will serve as input for the calculation of the pulsar contribution to the 
diffuse galactic \gr emission.

For each source we can define a critical distance, up to which the source
can be detected directly, and beyond which it would contribute to the
diffuse emission. We will assume that the Galaxy is a simple disk-like entity
of radius $r_h$ and half-thickness $z_h$, in which
the line-of-sight through the disk determines the sensitivity
threshold $F_c (l,b)$.
The numbers $r_h$, $z_h$ have
been chosen such that on average the threshold corresponds to the flux 
of a 5$\sigma$ source, respectively 4$\sigma$ for $\vert b\vert >$10$^\circ$,
in the summed data of Phases 1-4. The threshold
in direction of the galactic poles would
be $F_c (0,90) = 7\cdot 10^{-8}\,{\rm ph./cm^2 /sec}$, whereas in direction of the galactic center it
would be $F_c (0,0) = 4.8\cdot 10^{-7}\,{\rm ph./cm^2 /sec}$.
Then the critical distance can to be calculated as
\begin{equation}
X_{max} (l,b,t) = F_c^{-0.5} (l,b)\ \sqrt{\sum\limits_{k=3}^9\, S_m(t,E_k)\, \delta E_k}  \qquad {\rm kpc}\ \end{equation}
where $\delta E_k$ is the width of the energy bin $E_k$.
\begin{figure}
\centerline{\epsfig{file=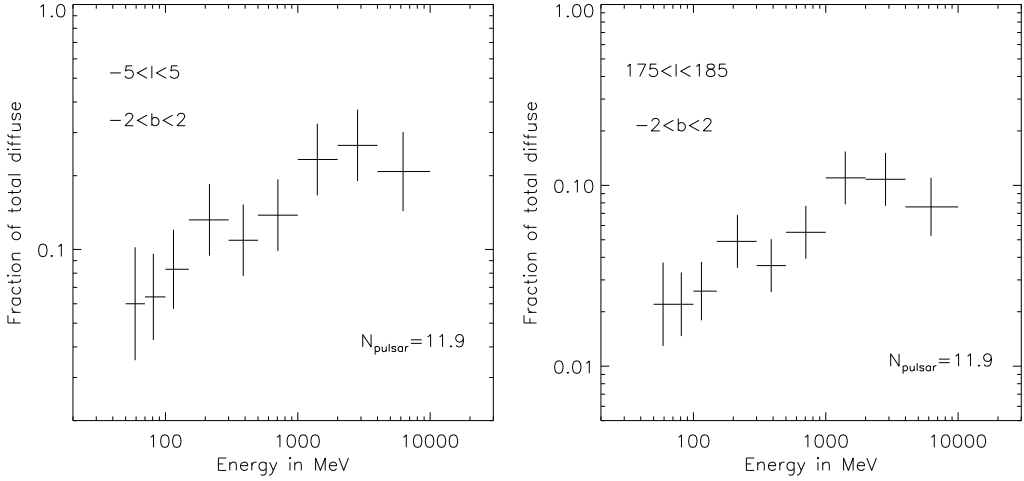}}
\caption{Here we show the fraction of the total diffuse intensity in direction
of the Galactic Center (left) and galactic anticenter (right)
which can be attributed to unresolved 
pulsars. The parameters to the model are given in the text. 
The error bars are derived by propagation of the parameter uncertainty
in the pulsar model fit.
The intensity due to pulsars scales almost linearly with the number of objects
which are supposed to be seen as point sources, in this case 
11.9 pulsars.}
\label{grad}
\end{figure}
The spatial distribution of pulsars is poorly determined. The distribution
of radio pulsars indicates that there is a galactocentric gradient
\cite{tayl93}.
We may thus parametrize the normalized spatial distribution of pulsars
in galactocentric cylinder coordinates (r,z) as
\begin{equation}
\rho (r,z) =0.0435\ \Theta(\mid z_c -z\mid )\ (z_c\, r_c^2)^{-1}\ \cosh^{-1}
\left({{r}\over {r_c}}\right)\ \end{equation}
where the $\cosh$-term accounts for the galactocentric gradient
and $\Theta$ is a step function.
With $\tau_p^{-1}$ as the birth rate of
\gr pulsars and $\epsilon $ as the fraction which radiates in
our direction we can determine the number of directly detectable pulsars 
by integration over the line-of-sight, solid angle, and age
\begin{equation}
N_{det} = {{\epsilon}\over {\tau_p}} \int_0^{t_{max}} dt\ 
\oint d\Omega\ \int_0^{X_{max}} dx\ x^2\,\rho (r,z)
\end{equation}
as well as the diffuse emission of unresolved pulsars
\begin{equation}
I_{dif} (E_k) = {{\epsilon}\over {\tau_p\,\Omega}} \int_0^{t_{max}} dt\ 
\int_{\Omega} d\Omega^\prime\ \int_{X_{max}}^\infty dx\ \rho (r,z)\, S_m(t,E_k)
\end{equation}
where $x$ is the distance coordinate along the line-of-sight.
Please note that any change in the total number of pulsars has similar
impact on the number of directly observable pulsars as on the \gr intensity
of unresolved objects. Thus the direct detections provide a strong constraint
on the pulsar contribution to the diffuse galactic \gr emission and limit
our choice of $\epsilon/\tau_p$.
\begin{figure}
\centerline{\epsfig{file=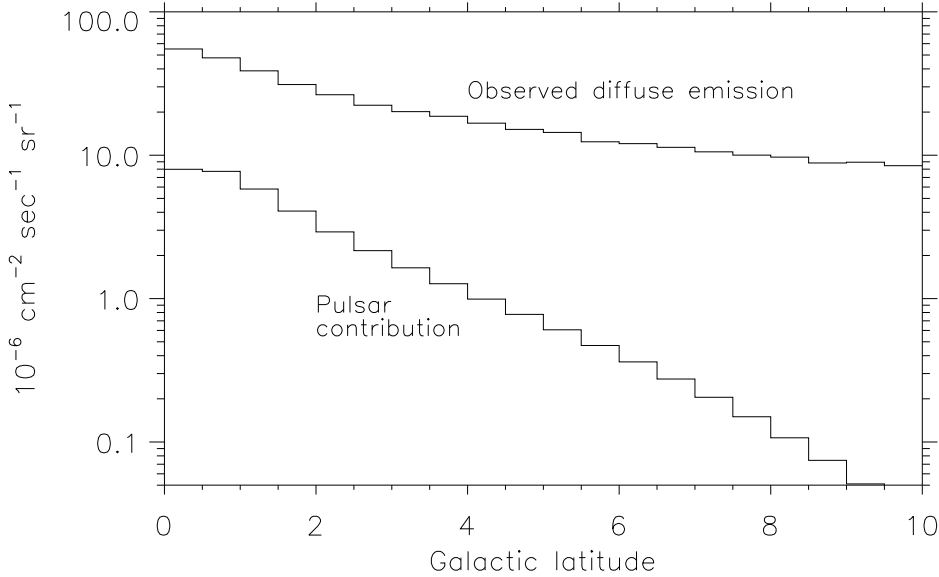}}
\caption{The latitude distribution of the diffuse emission
above 1 GeV as observed by EGRET is much wider than that of the pulsar
contribution, here shown as solid line for 12 directly observable pulsars.
Both the data and the model have been averaged over longitude. Here
we show the original observed intensity
distribution compared to the psf-convolved model prediction. The statistical 
error of the observed emission is at the percent level and thus negligible.}
\label{dif}
\end{figure}
With Eq.5 we can now calculate the spectrum of unresolved pulsars and
compare it to the spectrum of the total observed diffuse emission. This is
shown for the Galactic Center direction and for the anticenter in
Fig.\ref{grad}, where we plot the fraction of the total observed emission
which is due to pulsars.
Here we used for the spatial distribution of pulsars
a radial scale length $r_c$=3.5 kpc and a half-thickness of $z_c$=0.15 kpc.
The result changes little when we choose an age-dependent thickness
which would account for older pulsars being more distant from the plane.
The maximum age $t_{max}$, we have used $2\cdot 10^6\,$years, has
only little influence on the result since it affects the number of 
directly observable objects $N_{det}$ nearly as much as the predicted 
emission of unresolved pulsars $I_{dif}$.

The parameters used here ($\tau_p^{-1} = 0.01\ {\rm years^{-1}}$
and $\epsilon = 0.15$) imply that around 12 pulsars should be detectable
by EGRET. 
Since six are already identified this would mean that another
six unidentified sources are actually pulsars. 
The small fraction of unresolved EGRET sources which shows pulsar-like
or Geminga-like \gr spectra \cite{merc96} argues strongly against
the bulk of unidentified sources being pulsars. Also a substantial
fraction of the unidentified EGRET sources in the galactic plane
appears to be variable which makes the identification of {\it all}
unidentified sources with pulsars even more unlikely \cite{mcla96}.
In total, we think that around 12 directly observable pulsars, of which six
are already identified, is a reasonable number, and that thus the calculated
contribution of unresolved pulsars can be taken as serious estimate. 
Considering the integrated
intensity above 100 MeV in this example pulsars would cause 13\% of the 
observed intensity in the Galactic Center direction and about 5\% in the
anticenter direction. We have compared the latitude distribution of the
observed diffuse emission above 1 GeV
to what our model predicts for the pulsar contribution.
The result is given in Fig.\ref{dif} where we show the distribution of the diffuse
intensity at energies above 1 GeV as averages over longitude.
We find that pulsars contribute only very close to the plane where
the line-of-sights are long, but the observed emission does fall off
much less rapidly with latitude than the emission of unresolved pulsars.
Also the observed spectral discrepancy above 1 GeV seems
to extend to latitudes $|b|\ge 5^{\circ}$.
Thus there must be
additional effects playing a r\^ ole for the observed diffuse \gr emission
above 1 GeV.

\vskip0.2truecm
\noindent
The EGRET Team gratefully acknowledges support from the following:
Bundesministerium f\"{u}r Bildung, Wissenschaft, Forschung und Technologie, 
Grant 50 QV 9095 (MPE); and NASA Cooperative Agreement NCC 5-95 (SU).

\end{document}